\documentclass[pdflatex,sn-mathphys]{sn-jnl}

\jyear{2021}%
\usepackage{enumitem}
\theoremstyle{thmstyleone}%
\theoremstyle{thmstyletwo}%

\theoremstyle{thmstylethree}%

\begin{document}

\title[Monu yadav and Laxminarayan Das]{Evaluation of the Radius of Maximum Wind over the North Indian Basin with the help of Tropical Cyclone characteristics }


\author*[1]{\fnm{Monu} \sur{Yadav}}\email{yadavm012@gmail.com}
\author[1]{\fnm{Laxminarayan} \sur{Das}}\email{lndas@dtu.ac.in}



\affil*[1]{\orgdiv{Department of Applied Mathematics}, \orgname{Delhi Technological University}, \orgaddress{ \city{New Delhi}, \country{India}}}




\abstract{Tropical Cyclones (TCs) have devastating effects on several coastal regions worldwide. Precautionary knowledge about TC characteristics such as wind direction, wind speed, epicenter position, condensed vapor pressure measure, and radius of maximum can be highly valuable in disaster management and economic planning. Existing literature has focused on TC wind direction, intensity, cloud shape, and epicenter position, but there has been limited research on estimation of the Radius of Maximum Wind (RMW). Accurate estimation of RMW is crucial as errors can significantly impact wind and storm surge assessments and forecasts.
	
In this study, our objective is to determine the RMW over the North Indian Ocean (NIO). We chose this region due to its location surrounded by the Bay of Bengal and Arabian Sea, making it one of the six globally prominent areas prone to TCs. Our study is on the relationship between the center of the TC, the estimated pressure drop at the center, and the RMW, using historical observations and mathematical correlations.
	
To address missing parameters in the best track database of the Indian Meteorological Department, we employ a local regression model. We validate the accuracy of our developed method using two statistical measures: error percentage and T-test. Numerous TC cases are discussed in the paper over the NIO. Our findings indicates that the suggested method exhibits an error percentage ranging from approximately $-63\%$ to $50\%$ when compared to the best track data provided by the Indian Meteorological Department (IMD). In contrast, the error percentages for two other references \cite{bib21, bib22} with the same best track data range from approximately $-26\%$ to $200\%$. Moreover, the T-test results demonstrate that our method outperforms than the other approaches in terms of statistical significance. 
 }




\maketitle

\section{Introduction}\label{sec1}
Tropical Cyclones (TCs) are considered the most destructive natural disasters, and their potential causes damage that influences by their lifespan and size \cite{bib1}. The devastating effects of these cyclones are primarily caused by strong winds, heavy rainfall, and storm surges. A TC refers to a large-scale, non-frontal low-pressure system with a warm core that develops over tropical or sub-tropical waters. When cyclonic disturbances exceed a minimum threshold of 34 knots, they are universally classified as ``Tropical Cyclones" by the World Meteorological Organization. Annually, around 90 TCs are observed worldwide \cite{bib2}. Among the components of a TC, strong winds have a significant impact on property damage and secondary risks like storm surges and increased ocean wave functionality.

Despite meteorologists and warning centers studying winds associated with TCs since the 19th century, there is still room for improvement in understanding and predicting these winds functionality. Accurate wind information is crucial for minimizing the human and economic losses caused by cyclones. One important wind-related parameter is the RMW, which represents the distance between the cyclone's center and its band of strongest winds. The RMW varies across cyclones and can greatly influence the societal impacts of storms with similar intensity and size. The errors in RMW estimation can affect wind and storm surge assessments and forecasts.

While numerous studies have focused on predicting and estimating various aspects of TCs, including intensity, temperature, moisture precipitation, pressure systems, and cloud shapes, only a few studies have centered on the RMW specifically over the Northwest Pacific (NP), North Indian (NOI), and South Indian (SOI) basins. We have drawm more attention to this parameter, as accurately determining the location and extent of storm hazards, such as extreme wind circulation, coastal rainfall, and storm surge, can aid governments in taking necessary steps for disaster prevention and mitigation.

Various researchers have proposed models and equations for estimating the RMW using different datasets and approaches. Here are few of them like as:

\begin{enumerate} [label={(\alph*)}]
	\item The Monte Carlo (MC) wind probability model, that developed by Gross et al. \cite{bib28}, incorporates error characteristics from official track and intensity forecasts. It also takes into account climatological variations in TC size of Atlantic storms generated west of 55°W latitute between 1988 and 2002. In contrast to other methods which describing wind probability, it directly samples the error distributions, eliminating the need to fit assumed functional forms to the errors. Additionally, the MC method considers serial correlations.
	
	Gross et al. \cite{bib28} established relationships between the RMW ($r_{max}$) (in nautical miles), maximum wind speed $V_{max}$, and latitude.
	\begin{equation}
		r_{max} = 35.37 - 0.111 \times V_{max} + 0.570(\theta - 25)
	\end{equation}
	where $\theta$ is latitude (degrees), the maximum wind speed (kt). 
	
	\item Willoughby et al. \cite{bib21} introduced an empirically derived model that describes the structure of the hurricane vortex. This model serves as a valuable tool for various purposes, including theoretical studies on vortex dynamics, storm-surge prediction, and windstorm loss estimation. The model's estimates $r_{max}$ is:
	\begin{equation}
		r_{max} = 46.6 \times e^{(-0.015V_{max} + 0.0169 \theta)}
		\label{eq2}
	\end{equation}

    \item Lajoie et al. \cite{bib32} have developed a simple technique that utilizes the characteristic cloud and wind structure of a TC's eyewall. This technique demonstrates that the RMW depends on both the radius of the eye and the distance from the center to the top of the most developed cumulonimbus cloud closest to the cyclone center. In order to determine these two parameters, the proposed technique involves analyzing high-resolution IR and microwave satellite imagery. To evaluate the effectiveness of this technique, the derived RMW was compared with high-resolution wind analyses compiled by the U.S. National Hurricane Center and the Atlantic Oceanographic and Meteorological Laboratory. The relation by Lajoie et al. is:
    \begin{equation}
    	r_{max} =  \left(1-\frac{h_v}{h_t}\right) \times r_t + \left(\frac{h_v}{h_t}\right) \times r_e
    \end{equation}
    Here, $r_e$ is eye radius, $r_t$ is the distance between the cyclone center and the cloud-top temperature nearest to the cyclone center. 
    
    \item Lajoie et al. \cite{bib32} have developed a simple technique that utilizes the characteristic cloud and wind structure of a TC's eyewall. This technique demonstrates that the RMW depends on both the radius of the eye and the distance from the center to the top of the most developed cumulonimbus cloud closest to the cyclone center. In order to determine these two parameters, the proposed technique involves analyzing high-resolution IR and microwave satellite imagery. To evaluate the effectiveness of this technique, the derived RMW was compared with high-resolution wind analyses compiled by the U.S. National Hurricane Center and the Atlantic Oceanographic and Meteorological Laboratory. The relation by Lajoie et al. is:
    \begin{equation}
    	r_{max} =  \left(1-\frac{h_v}{h_t}\right) \times r_t + \left(\frac{h_v}{h_t}\right) \times r_e
    \end{equation}
    Here, $r_e$ is eye radius, $r_t$ is the distance between the cyclone center and the cloud-top temperature nearest to the cyclone center. 
    
    \item Vickery and Wadhera \cite{bib29} conducted a study where they utilized the HWind (HRD Real-time Hurricane Wind Analysis System) data and an upper-level aircraft dataset to develop statistical models for the Holland B parameter and $r_{max}$. The HWind system functions as a distributed system that collects real-time observations of TCs from various platforms such as land, sea, space, and air. These observations are then standardized and presented graphically in relation to the storm, providing interactive tools for scientists to analyze, validate, and visualize the information.
    
    To ensure the reliability of their models, the authors focused on a dataset comprising 2,291 radial profiles from 62 Atlantic TCs. They specifically selected storms with central pressures below 980 hPa for their analysis. The study yielded multiple relations of the RMWs, with particular emphasis on hurricanes that made landfall and those that traveled over the Atlantic Ocean and Gulf of Mexico. The relationship they established between $r_{max}$ (in km), pressure drop, and latitude is as follows:
    \begin{equation}
    	ln(r_{max}) =  3.015- 6.291 \times 10^{-5} (\nabla p)^2 +0.337 \times \theta
    \end{equation}
    where $\nabla p$ is the pressure drop
    
    \item Quiring et al. \cite{bib31} studies provide an explanation about the average size of TC year-to-year in the Atlantic region. The relation derived from their model, representing $r_{max}$ across the entire Atlantic, has expressed as follows:
    \begin{equation}
    	r_{max} = 49.67 - 0.24 \times V_{max}
    \end{equation}

\item Takagi and Wu \cite{bib30} examined existing models used for estimating $r_{max}$ and put forward a new formula aimed at minimizing estimated errors specifically for storm surge modeling purposes. To ensure the integrity of their analysis, they utilized a dataset comprising observations from 17 TCs that had passed near 10 JMA stations situated in the southern islands of Japan. By utilizing the data from these remote islands, they were able to avoid the considerable alterations in typhoon structure induced by land topography. Moreover, they limited their dataset to typhoon central pressures below 935 hPa. Their proposed approach for estimating $r_{max}$ relied on the radius of 50-knot winds ($r_{50}$) measured in nautical miles.
\begin{equation}
	r_{max} = 0.23 \times r_{50}
\end{equation}
where $r_{50}$ is the radius of 50-kt winds (nmi). They further recommended that to minimize over- or underestimation of storm surges, simulations should be repeated for different estimations of $r_{max}$ covering a percentage of the data (e.g. $r_{max}$ = $0.15 r_{50}$ to $0.35r_{50}$).

\item Tan and Fang \cite{bib22} focused on simulating wind fields of historical TCs using parametric models using the satellite data. These models explicitly incorporated local effects at a resolution of 1 km, specifically considering the planetary boundary layer. The authors quantified the topographic effects for eight wind directions across four terrain types: ground, escarpment, ridge, and valley. To estimate surface roughness lengths, they utilized a global land cover map.

To address missing TC parameters in the best track datasets, the authors employed local regression models for reconstruction. Their study is also encompassed a global wind hazard assessment, with a particular emphasis on the NIO. Notably, the results revealed a significant absence of parameters, amounting to approximately $68.56\%$. The relation of $r_{max}$ is: 
\begin{equation}
	r_{max} = -26.73 \times ln(1013.25 - P_c) + 142.41 
	\label{eq3}
\end{equation}
where $P_c$ is the estimated central pressure of TC

\item Chavas et al. \cite{bib27} employed the principles of angular momentum loss within the inflow of TCs to establish a connection between the two radii, dependent on two easily defined physical parameters. To estimate the model coefficients, they make use of observational data. The relation formulated by Chavas et al. is:
\begin{equation}
	r_{max} = \frac{V_{max}}{f} \times \sqrt{1+ \frac{2f M_{max}}{V_{max}^2} - 1}
\end{equation}
where f is the Coriolis parameter, M is the angular momentum.
\end{enumerate}

These method are performed over the either particular basin or group of two or three basins. Some countries also have developed methods at national-level like the HAZUS hurricane model in the United States has undergone continuous refinement since 1997 \cite{bib11}. In Florida, the Florida Public Hurricane Loss Model has been established with support from the Florida Office of Insurance Regulation \cite{bib20}. Central American nations have implemented the Comprehensive Approach to Probabilistic Risk Assessment (CAPRA) \cite{bib12}, while Australian has its TC Risk Model (TCRM) \cite{bib13,bib14}. Furthermore, a global parametric wind circulation field model, known as the Holland model, was used to recreate historical TC wind fields globally for all basins from 1970 to 2010 \cite{bib15,bib23,bib24}. Subsequent enhancements and improvements were made to this methodology in 2013 and 2015 \cite{bib16,bib25,bib26}. However, these models still suffer from significant errors. Improving the accuracy of the RMW estimation is crucial for accurately predicting other cyclone characteristics like intensity, direction of wind, and pressure. It can also help assess trends in human and economic losses, enabling governments to take necessary measures.

Keeping the above fact in mind we specifically focuses on the NIO, surrounded by the Bay of Bengal and the Arabian Sea. The Bay of Bengal is recognized as one of the six prominent regions globally prone to TCs, with a significant number of devastating cyclones originating from this area. Historically, these cyclones have mostly occurred between October and November. A concerning trend in the Bay of Bengal is the rising sea temperatures, which contribute to the intensification of cyclones as they approach coastal regions. The increased sea surface temperatures create a favorable environment for cyclones to strengthen, amplifying the associated risks and impacts.

Our work aims to address the existing knowledge gap by achieving the following objectives:
\begin{enumerate}
	\item developing a simple estimation model for the RMW using key TC parameters
	\item  estimating model parameters based on historical data from the NIO basin
	\item evaluating the model associated observational records
	\item comparing the model's performance with existing predictive models for the RMW over the NIO.
\end{enumerate}  

This paper is organized as follows: Section \ref{sec2} presents the data and methodology used in the research. The proposed method is discussed in detail in Section \ref{sec3} and compared with two other approaches mentioned in the referenced papers \cite{bib21,bib22}. Section \ref{sec4} provides the study's conclusion, while Section \ref{sec5} outlines potential avenues for future research.

\section{Data and Methodology}\label{sec2}
This study evaluating a suggested method using the best track dataset obtained from the Indian Meteorological Department (IMD), which serves as the Regional Specialized Meteorological Center for TCs over the NIO. The dataset consists of various metrics related to TCs, including the TC number, time (year, month, day, and hour), TC center locations (longitude and latitude), estimated central pressure ($P_{c}$), maximum wind speed ($V_{m}$), and estimated pressure drop at the center ($P_{d}$). These metrics are recorded at 3-hour intervals. We use local regression models for the missing parameter in the best track database. To validate the suggested method, we also compared it with the RMW data from IMD bulletins.

The main objective of this study was to develop a method for calculating the RMW ($r_{max}$) based on the characteristics of TCs, such as the latitude (Lat.) and longitude (Long.) of the TC center, along with the estimated pressure drop at the TC center. Through an analysis of relevant variables and their impact on the RMW, a relationship was established using historical observations over the NIO with mathematical correlation. This relationship allows us to determine the value of $r_{max}$ for estimated pressure drops ($P_d$) that are less than or equal to 12 hPa.

The relationship for determining $r_{max}$ (in nautical miles) is as follows:
\begin{equation}
	r_{max} = 10.16812 \times e^{(-0.213\sqrt{P_d} + 0.0169\phi)} + 23.817
\end{equation}
Here, $\phi$ represents the latitude of the TC center, and $P_d$ represents the estimated pressure drop at the TC center.

This method provides a reliable means of calculating the RMW based on the given TC characteristics. The relationship established in this study allows for accurate determination of $r_{max}$ as long as the value of $P_d$ is less than or equal to 12 hPa. This information can contribute to better understanding and prediction of TCs characteristics, aiding in mitigation efforts and preparedness for such weather events.

The proposed method for determining the value of $r_{max}$ was evaluated using seven TC cases, namely: Extremely Severe Cyclonic Storm Tauktae, Cyclonic Storm Gulab, Severe Cyclonic Storm Mandous, Severe Cyclonic Storm Asani, Severe Cyclonic Sitrang, Cyclonic Storm Jawad, and Very Severe Cyclonic Storm Yaas. These specific cases were chosen to accurately assess the suggested method's effectiveness.

To validate the accuracy of the proposed method, two evaluation techniques were employed: error percentage and statistical T-test. By utilizing these methods, we can confidently assert that the suggested approach outperforms alternative methods used for determining $r_{max}$ in the NIO region. We compare our designed method with the models proposed by Willoughby et al. \cite{bib21} and Tan and Fang \cite{bib22}, both specifically designed for NIO cases.

Firstly, we calculated the error percentage by comparing the values obtained from the suggested method with the data provided by the IMD and the results of the other two methods \cite{bib21,bib22}. The error percentage formula used is as follows:

\begin{equation*}
	\text{Error percentage} = \frac{\text{Experimental Value} - \text{Actual Value}}{\text{Actual Value} }\times 100
\end{equation*}

Once the error percentages were calculated, a statistical T-test was conducted to determine the accuracy of the suggested method in comparison to the alternative approaches. The t-test is a widely used statistical hypothesis test that enables a comparison of means between two groups, thereby providing valuable insights into method's accuracy. For this analysis, we employed the Python language along with the numpy and scipy libraries to calculate the T-tests effectively.

\section{Discussion}\label{sec3}

\subsection{Discussion on Extremely Severe Cyclonic Storm Tauktae}
The suggested method, along with the approaches proposed by Willoughby et al., Tan and Fang, and data provided by the IMD, is visually depicted in Figure 1. This graphical representation showcases the value of $r_{max}$. Table 1 complements this with a numerical summary of the results for the Extremely Severe Cyclonic Storm Tauktae. It is noteworthy that the suggested method exhibits error percentages ranging from $-9\%$ to $-50\%$, while the error percentages for the expressions by Willoughby et al. \cite{bib21} and Tan and Fang \cite{bib22} span from $-3\%$ to 66$\%$.

Upon conducting a T-test, it is evident that the suggested method shows statistically better over the other two approaches at $\alpha = 0.05$.

\begin{table}[h]\label{tab1}
	\begin{tabular}{p{25pt}|p{20pt}|p{40pt}|p{40pt}|p{30pt}|p{25pt}|p{25pt}|p{25pt}}
		\hline
		Date/ Time& IMD's $r_{max}$ &Suggested $r_{max}$&Willoughby et al. $r_{max}$ & Tan and Fang $r_{max}$ & $E_1$ ($\%$)& $E_2$ ($\%$)& $E_3$ ($\%$) \\
		\hline
		14/06&40&31.8147&38.571&66.288&-20.46&-3.57&65.72  \\
		15/00&60&30.7164&31.69&60.7136&-48.80& -47.18&1.18\\
		15/06&60&30.2973&29.65&58.30 &-49.50&-50.58&-2.83\\
		15/12&32&29.4437&25.785&53.10& -7.98&-19.42&65.93\\
		15/18&32&29.1063&24.207&50.4 &-9.04&-24.35&57.5\\
		\hline
	\end{tabular}

\caption{Results of Extremely Severe Cyclonic Storm Tauktae, where $E_1$ indicates the error percentage between suggested method and IMD, $E_2$ represents the error percentage between Willoughby et al.'s expression and IMD, and $E_3$ represents the error percentage between Tan and Fang's expression and IMD}
\end{table}
\begin{figure}\label{f1}
	\includegraphics[width=1\textwidth]{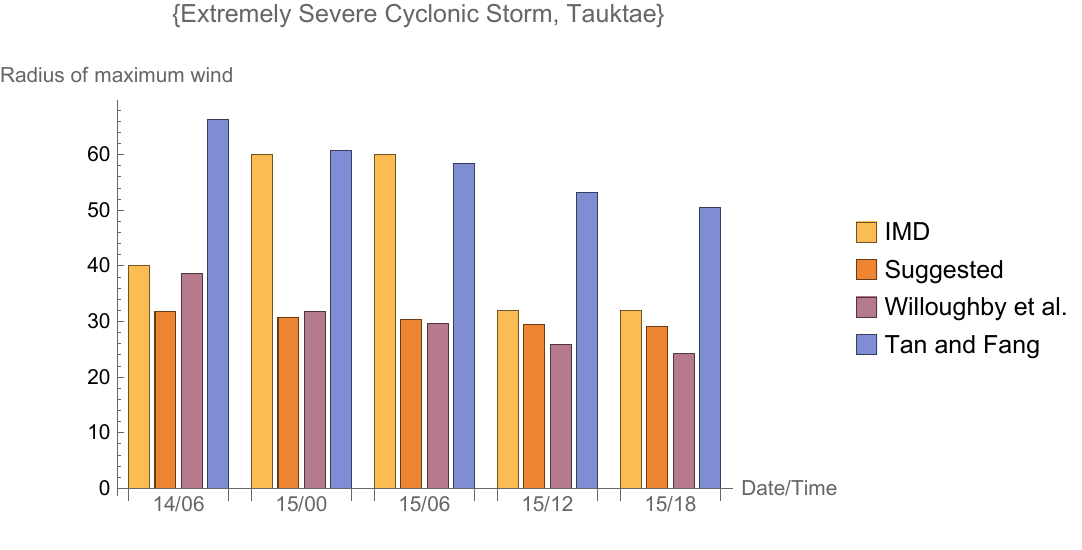}
	\caption{Graphical Representation of Extremely Severe Cyclonic Storm Tauktae's value of $r_{max}$ by suggested method, Willoughby et al., Tan and Fang, and IMD}
\end{figure}
\subsection{Discussion of Cyclonic Storm Gulab}
The value of $r_{max}$ as represented graphically by the suggested method, Willoughby et al., Tan and Fang, and IMD is shown in Figure 2. Table \ref{tab2} summarizes the Cyclonic Storm Gulab results numerically. We find that the error precentage of the suggested method ranges from $-42 \%$ to $51\%$, while that of the expressions of Willoughby et al. \cite{bib21} , and Tan and Fang \cite{bib22} ranges from $-26 \%$ to $202\%$.

The T-test reveals that the suggested method shows statistically better over the other two approaches at $\alpha = 0.05$.

\begin{table}[h]
	\begin{tabular}{p{25pt}|p{20pt}|p{40pt}|p{40pt}|p{30pt}|p{25pt}|p{25pt}|p{25pt}}
		\hline
	Date/ Time& IMD's $r_{max}$ &Suggested $r_{max}$&Willoughby et al. $r_{max}$ & Tan and Fang $r_{max}$ & $E_1$ ($\%$)& $E_2$ ($\%$)& $E_3$ ($\%$) \\
	\hline
	25/00&55&32.43&40.55&71.39&-41.02&-26.27&29.8 \\
	25/06&55&32.05&40.55&71.39&-41.72&-26.27&29.8\\
	25/12&24&31.70&37.55&65.28&32.09&56.45&174.5\\
	25/18&24&31.70&37.55&66.28&32.09&56.45&176.16\\
	26/00&30&31.40&34.84&63.35&4.67&16.13&111.16\\
	26/06&24&30.89&32.38&60.71&28.71&34.91&152.95\\
	26/12&24&30.89&32.38&60.71&28.71&34.91&152.95\\
	26/18&21&31.71&37.62&63.35&51.02&79.14&201.66\\
	\hline
	\end{tabular}
\caption{Results of Cyclonic Storm Gulab, where $E_1$ indicates the error percentage between suggested method and IMD, $E_2$ represents the error percentage between Willoughby et al.'s expression  and IMD, and $E_3$ represents the error percentage between Tan and Fang's expression and IMD}
\label{tab2}
\end{table}
\begin{figure}\label{fig2}
	\includegraphics[width=1 \textwidth]{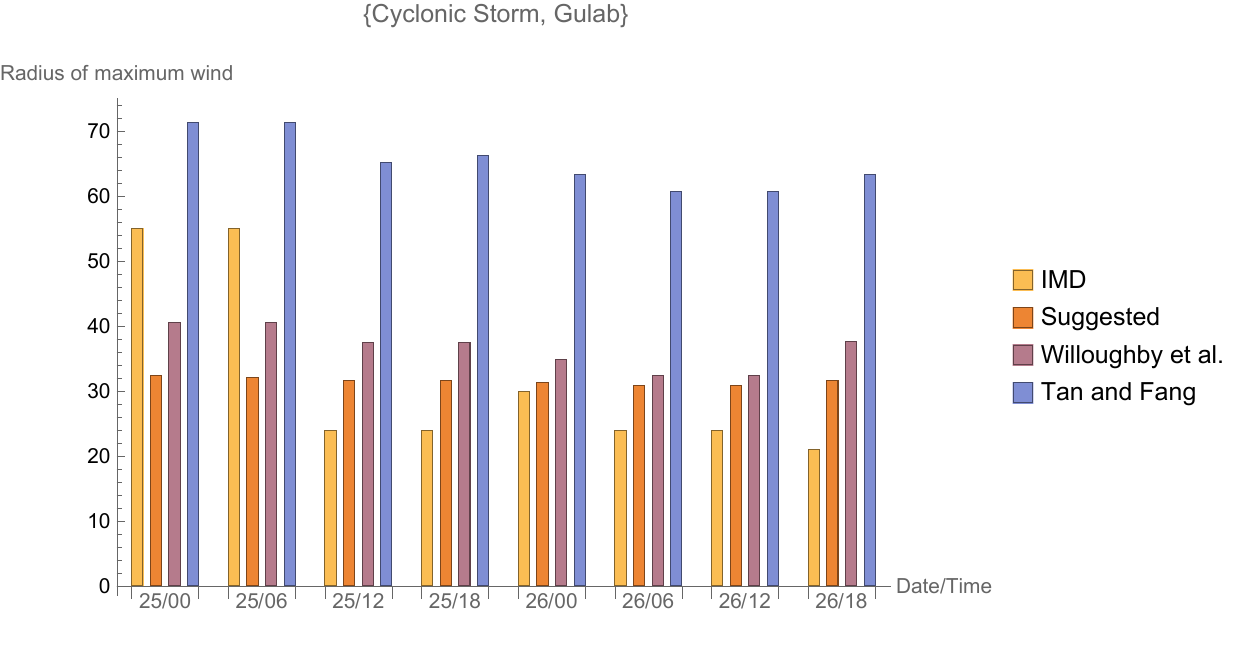}
	\caption{ Cyclonic Storm Gulab's value of $r_{max}$ given by suggested method, Willoughby et al., Tan and Fang, and IMD is graphically represented}
\end{figure}
\subsection{Discussion of Severe Cyclonic Storm Mandous}
The value of $r_{max}$ as represented graphically by the suggested method, Willoughby et al., Tan and Fang, and IMD is shown in Figure 3. Table \ref{tab3} summarizes the Severe Cyclonic Storm Mandous results numerically. We find that the error precentage of the suggested method ranges from $-52 \%$ to $35\%$, while that of the expressions of Willoughby et al. \cite{bib21} , and Tan and Fang \cite{bib22} ranges from $-47 \%$ to $165\%$.

The T-test reveals that the suggested method shows statistically better over the other two approaches at $\alpha = 0.10$.

\begin{table*}[h]
	\begin{tabular}{p{25pt}|p{20pt}|p{40pt}|p{45pt}|p{30pt}|p{30pt}|p{30pt}|p{30pt}}
		\hline
		Date/ Time& IMD $r_{max}$ &Suggested $r_{max}$ & Willoughby et al. $r_{max}$ & Tan and Fang $r_{max}$ & $E_1$ ($\%$)& $E_2$ ($\%$)& $E_3$ ($\%$) \\
		\hline
		07/00 & 64 & 31.12 & 34.36 & 71.39 & -51.37 & -46.31 & 11.54 \\
		07/06 & 64 & 31.13 & 34.41 & 69.58 & -51.35 & -46.23 & 8.71 \\
		07/12 & 64 & 30.83 & 34.53 & 69.58 & -51.82 & -46.04 & 8.71 \\
		07/18 & 34 & 30.57 & 32.203 & 67.88 & -10.08 & -5.28 & -76.82 \\
		08/00 & 34 & 30.34 & 29.97 & 64.78 & -10.76 & -11.85 & 90.52 \\
		08/06 & 34 & 29.92 & 27.95 & 60.71 & -12 & -17.79 & 78.55 \\
		08/12 & 22 & 29.58 & 26.15 & 58.309 & 34.45 & 18.86 & 165.04 \\
		08/18 & 22 & 29.63 & 26.33 & 58.309 & 34.68 & 19.68 & 165.04 \\
		09/00 & 22 & 29.67 & 26.509 & 58.309 & 34.86 & 20.49 & 165.04 \\
		09/06 & 34 & 30.103 & 28.76 & 60.71 & -11.46 & -15.41 & 78.55 \\
		09/12 & 34 & 30.39 & 31.32 & 62.002 & -10.61 & -7.88 & 82.35 \\
		09/18 & 34 & 30.96 & 34.05 & 64.781 & -8.94 & 0.14 & 90.53 \\
		10/00 & 64 & 31.67 & 36.95 & 69.58 & -50.51 & -42.26 & 8.71 \\
		\hline
	\end{tabular}
	\caption{ Results of Severe Cyclonic Storm MANDOUS over Bay of Bengal during $6-10$ December, 2022, where $E_1$ indicates the error percentage between suggested method and IMD, $E_2$ represents the error percentage between Willoughby et al.'s expression and IMD, and $E_3$ represents the error percentage between Tan and Fang's expression and IMD    }
		\label{tab3}
\end{table*}
\begin{figure}\label{fig3}
	\includegraphics[width=1\textwidth]{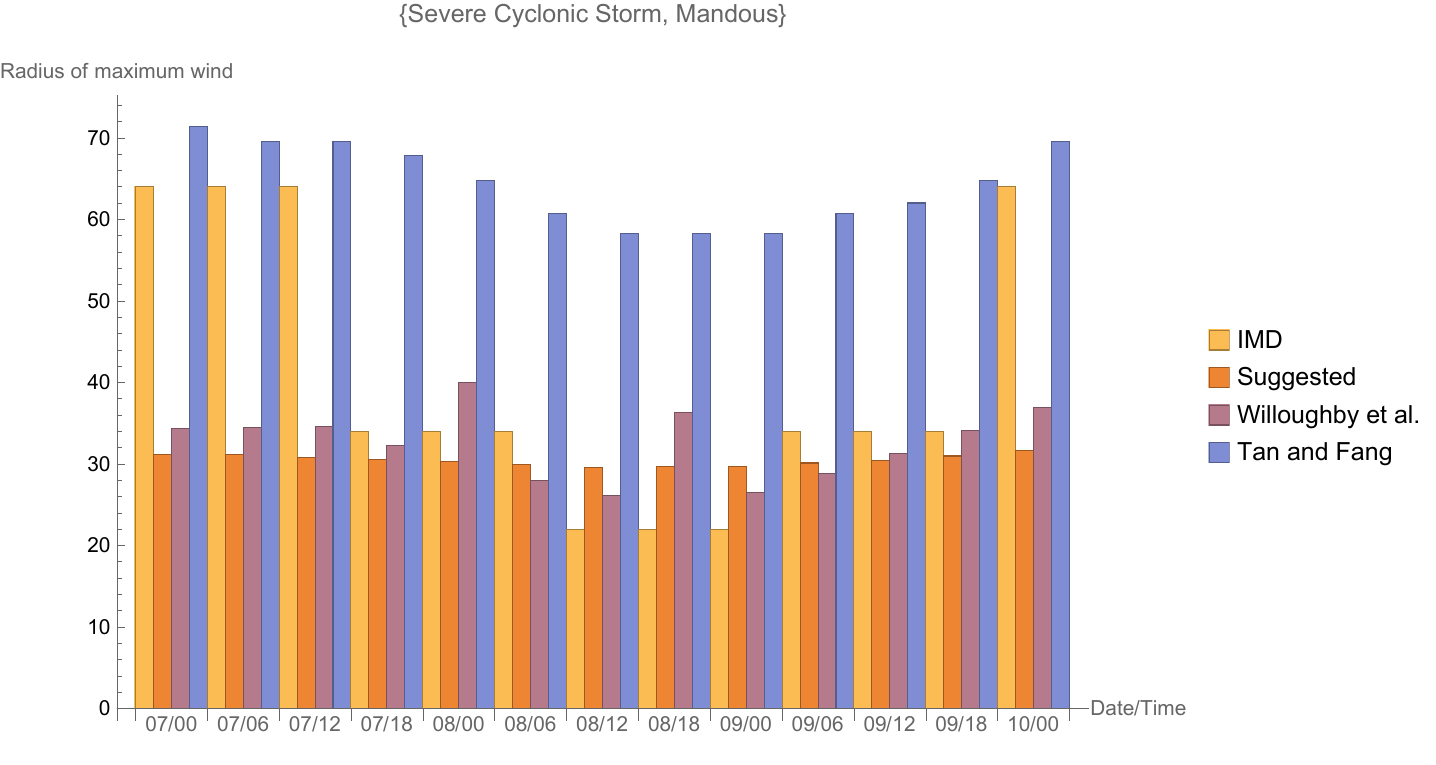}
	\caption{Severe Cyclonic Storm Mandous's value of $r_{max}$ given by suggested method, Willoughby et al., Tan and Fang, and IMD is graphically represented}
\end{figure}
\subsection{Discussion of Severe Cyclonic Storm Asani}
The value of $r_{max}$ as represented graphically by the suggested method, Willoughby et al., Tan and Fang, and IMD is shown in Figure 4. Table \ref{tab4} summarizes the Severe Cyclonic Storm Asani results numerically. We find that the error precentage of the suggested method ranges from $-42 \%$ to $4\%$, while that of the expressions of Willoughby et al.  \cite{bib21}, and Tan and Fang \cite{bib22} ranges from $-34 \%$ to $102\%$.

The T-test reveals that the suggested method shows statistically better over the other two approaches at $\alpha = 0.10$.

\begin{table*}
	\begin{tabular}{p{25pt}|p{20pt}|p{40pt}|p{45pt}|p{30pt}|p{30pt}|p{30pt}|p{30pt}}
		\hline
		Date/ Time& IMD $r_{max}$&Suggested $r_{max}$&Willoughby et al. $r_{max}$& Tan and Fang $r_{max}$& $E_1$ ($\%$)& $E_2$ ($\%$)& $E_3$ ($\%$) \\
		\hline
		07/12 & 55 & 31.33 & 35.303 & 64.78 & -43.03 & -35.81 & 17.78 \\
		07/18 & 55 & 31.39 & 35.66 & 63.35 & -42.91 & -35.16 & 15.18 \\
		08/00 & 30 & 30.81 & 33.31 & 60.71 & 2.70 & 11.03 & 102.6 \\
		08/06 & 30 & 30.12 & 28.86 & 56.10 & 0.41 & -3.8 & 87.01 \\
		11/00 & 30 & 30.58 & 30.98 & 56.10 & 1.96 & 3.26 & 87.01 \\
		11/06 & 30 & 31.11 & 33.51 & 58.30& 3.707 & 11.7 & 94.36 \\
		11/12 & 55 & 31.76 & 36.309 & 60.71 & -42.24 & -33.98 & 10.38 \\
		\hline
	\end{tabular}
	\caption{ Results of Severe Cyclonic storm, Asani over Bay of Bengal during $07-12$ May, 2022, where $E_1$ indicates the error percentage between suggested method and IMD, $E_2$ represents the error percentage between Willoughby et al.'s expression and IMD, and $E_3$ represents the error percentage between Tan and Fang's expression and IMD}
		\label{tab4}
\end{table*}

\begin{figure} \label{fig4}
	\includegraphics[width=1\textwidth]{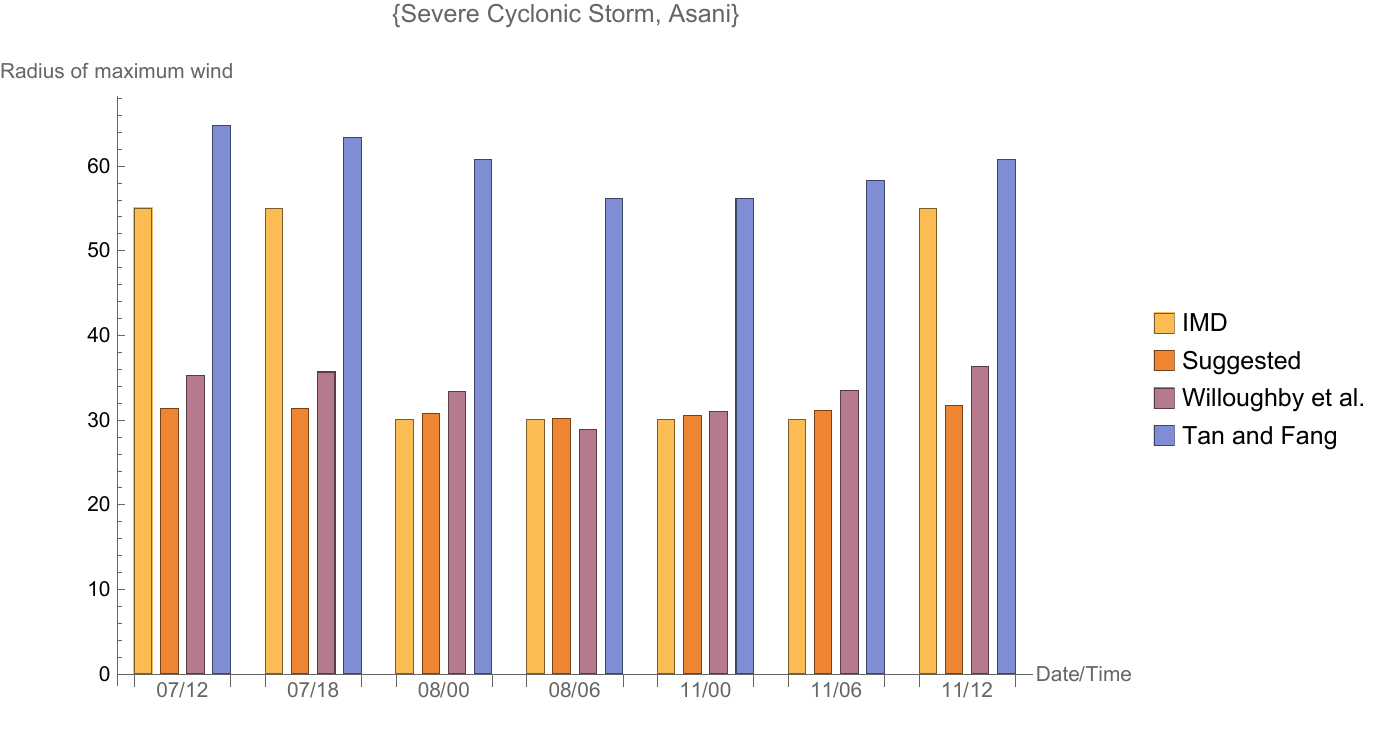}
	\caption{Severe Cyclonic Storm Asani's value of $r_{max}$ given by suggested method, Willoughby et al., Tan and Fang, and IMD is graphically represented}
\end{figure}
\subsection{Discussion of Cyclonic Storm Sitrang}
The value of $r_{max}$ as represented graphically by the suggested method, Willoughby et al., Tan and Fang, and IMD is shown in Figure 5. Table \ref{tab5} summarizes the Severe Cyclonic Storm Sitrang results numerically. We find that the error precentage of the suggested method ranges from $-5 \%$ to $-51\%$, while that of the expressions of Willoughby et al. \cite{bib21} , and Tan and Fang \cite{bib22} ranges from $-40 \%$ to $110\%$.

The T-test reveals that the suggested method shows statistically better over the other two approaches at $\alpha = 0.05$.

\begin{table*}[h]
	\begin{tabular}{p{25pt}|p{20pt}|p{40pt}|p{45pt}|p{30pt}|p{30pt}|p{30pt}|p{30pt}}
		\hline
		Date/ Time& IMD $r_{max}$ &Suggested $r_{max}$&Willoughby et al. $r_{max}$& Tan and Fang $r_{max}$& $E_1$ ($\%$)& $E_2$ ($\%$)& $E_3$ ($\%$) \\
		\hline
		23/00 & 64 & 31.99 & 38.48 & 75.43 & -50.00 & -39.87 & 17.85 \\
		23/06 & 64 & 31.71 & 38.87 & 73.33 & -50.45 & -39.26 & 14.57 \\
		23/12 & 34 & 31.49 & 36.55 & 71.39 & -7.37 & 7.5 & 109.97 \\
		23/18 & 34 & 31.57 & 36.92 & 69.58 & -7.14 & 8.58 & 104.64 \\
		24/00 & 34 & 31.09 & 34.66 & 66.28 & -8.55 & 1.94 & 94.94 \\
		24/06 & 34 & 31.001 & 32.87 & 66.28 & -8.82 & -3.32 & 94.94 \\
		24/12 & 34 & 31.14 & 33.55 & 66.28 & -8.38 & -1.32 & 94.94 \\
		24/18 & 34 & 32.09 & 35.29 & 66.28 & -5.59 & 3.79 & 94.94 \\
		\hline
	\end{tabular}
	\caption{ Results of Cyclonic storm, Sitrang over Bay of Bengal during $22-25$ October, 2022, where $E_1$ indicates the error percentage between suggested method and IMD, $E_2$ represents the error percentage between Willoughby et al.'s expression and IMD, and $E_3$ represents the error percentage between Tan and Fang's expression and IMD }
		\label{tab5}
\end{table*}
\begin{figure}[h]\label{fig5}
	\includegraphics[width=1\textwidth]{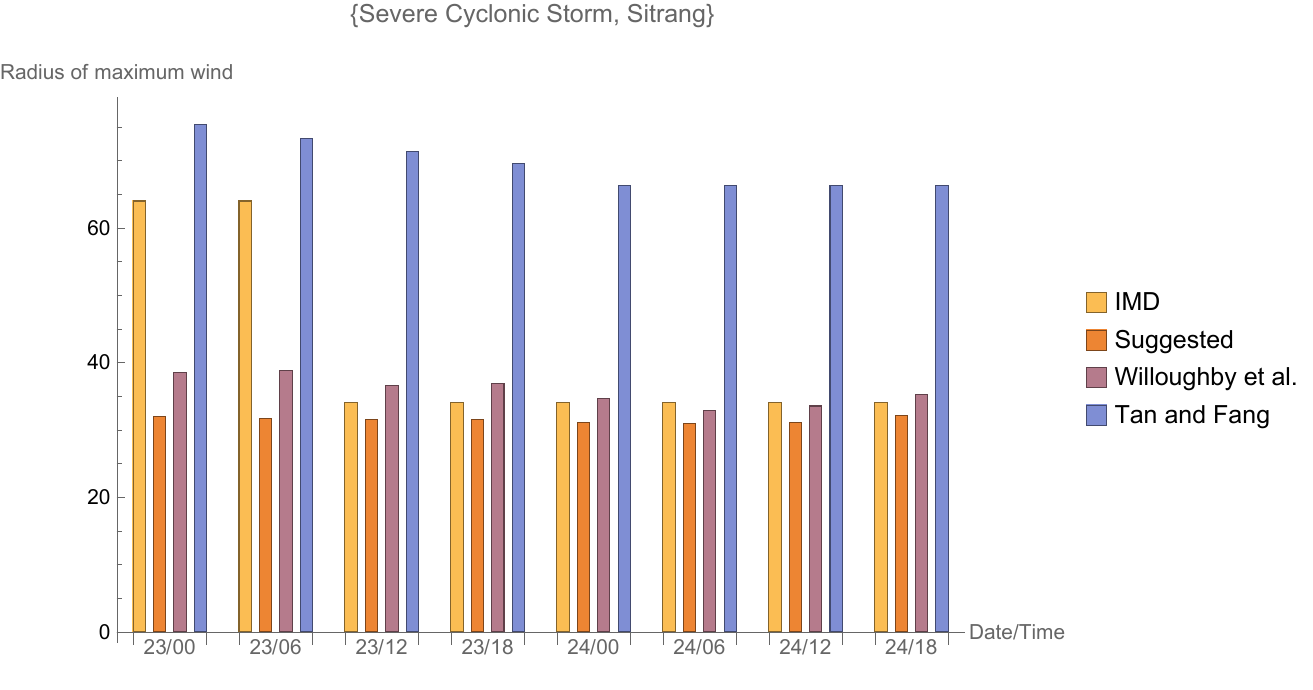}
	\caption{ Cyclonic Storm Sitrang's value of $r_{max}$ given by suggested method, Willoughby et al., Tan and Fang, and IMD is graphically represented}
\end{figure}
\subsection{Discussion of Cyclonic Storm Jawad}
The value of $r_{max}$ as represented graphically by the suggested method, Willoughby et al., Tan and Fang, and IMD is shown in Figure 6. Table \ref{tab6} summarizes the  Cyclonic Storm Jawad results numerically. We find that the error precentage of the suggested method ranges from $-25 \%$ to $-54\%$, while that of the expressions of Willoughby et al. \cite{bib21} , and Tan and Fang \cite{bib22} ranges from $-46 \%$ to $80\%$.

The T-test reveals that the suggested method shows statistically better over the other two approaches at $\alpha = 0.05$.

\begin{table*}[h]
	\begin{tabular}{p{25pt}|p{20pt}|p{40pt}|p{45pt}|p{30pt}|p{30pt}|p{30pt}|p{30pt}}
		\hline
		Date/ Time& IMD $r_{max}$&Suggested $r_{max}$&Willoughby et al. $r_{max}$& Tan and Fang $r_{max}$& $E_1$ ($\%$)& $E_2$ ($\%$)& $E_3$ ($\%$) \\
		\hline
		03/00 & 69 & 31.73 & 37.26 & 77.71 & -54.03 & -46 & 12.62 \\
		03/12 & 42 & 31.05 & 33.23 & 73.33 & -26.06 & -20.87 & 74.59 \\
		03/06 & 42 & 31.21 & 35.21 & 75.43 & -25.68 & -16.16 & 79.59 \\
		03/18 & 42 & 31.09 & 35.45 & 73.33 & -25.95 & -15.59 & 74.59 \\
		04/00 & 42 & 31.13 & 33.62 & 73.33 & -25.86 & -19.95 & 74.59 \\
		04/06 & 42 & 31.45 & 36.37 & 75.43 & -25.11 & -13.40 & 79.59 \\
		04/12 & 55 & 31.84 & 39.53 & 77.71 & -42.09 & -28.12 & 41.29 \\
		\hline
	\end{tabular}
	\caption{ Results of Cyclonic Storm, Jawad over the Bay of Bengal during $02-06$ December, 2021, where $E_1$ indicates the error percentage between suggested method and IMD, $E_2$ represents the error percentage between Willoughby et al.'s expression and IMD, and $E_3$ represents the error percentage between Tan and Fang's expression and IMD}
		\label{tab6}
\end{table*}
\begin{figure}[h]\label{fig6}
	\includegraphics[width=1\textwidth]{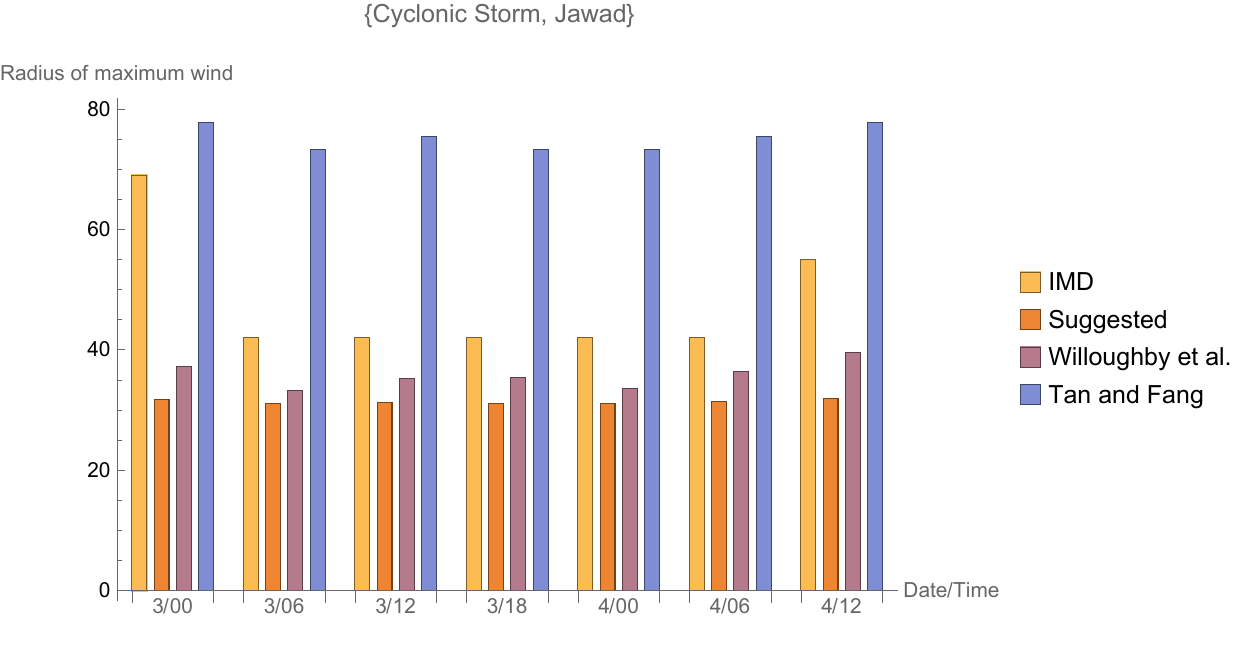}
	\caption{ Cyclonic Storm Jawad's value of $r_{max}$ given by suggested method, Willoughby et al., Tan and Fang, and IMD is graphically represented}
\end{figure}
\subsection{Discussion of Very Severe Cyclonic Storm Yaas}
The value of $r_{max}$ as represented graphically by the suggested method, Willoughby et al., Tan and Fang, and IMD is shown in Figure 7. Table \ref{tab7} summarizes the Very Severe Cyclonic Storm Yaas results numerically. We find that the error precentage of the suggested method ranges from $-20 \%$ to $-63\%$, while that of the expressions of Willoughby et al. \cite{bib21} , and Tan and Fang \cite{bib22} ranges from $-54 \%$ to $38\%$.

The T-test reveals that the suggested method shows statistically better over the other two approaches at $\alpha = 0.10$.

\begin{table*}[h]
	\begin{tabular}{p{25pt}|p{20pt}|p{40pt}|p{45pt}|p{30pt}|p{30pt}|p{30pt}|p{30pt}}
		\hline
		Date/ Time& IMD $r_{max}$ &Suggested $r_{max}$&Willoughby et al. $r_{max}$& Tan and Fang $r_{max}$& $E_1$ ($\%$)& $E_2$ ($\%$)& $E_3$ ($\%$) \\
		\hline
		23/18 & 85 & 32.13 & 39.13 & 60.71 & -62.19 & -53.96 & -28.57 \\
		24/00 & 58 & 31.44 & 36.309 & 58.309 & -45.79 & -37.39 & 0.53 \\
		24/06 & 58 & 31.16 & 33.74 & 26.103 & -46.27 & -41.82 & -3.27 \\
		24/12 & 58 & 30.73 & 31.67 & 54.06 & -47.00 & -45.39 & -6.79 \\
		24/18 & 38 & 30.36 & 29.63 & 52.17 & -20.09 & -22.02 & 37.28 \\
		25/00 & 38 & 29.69 & 27.68 & 48.74 & -21.85 & -27.15 & 28.26 \\
		25/06 & 38 & 29.46 & 25.98 & 47.18 & -22.45 & -31.63 & 24.15 \\
		\hline
	\end{tabular}
	\caption{ Results of  Very Severe Cyclonic Storm, “YAAS” over the Bay of Bengal during $23-28$ May, 2021, where $E_1$ indicates the error percentage between suggested method and IMD, $E_2$ represents the error percentage between Willoughby et al.'s expression and IMD, and $E_3$ represents the error percentage between Tan and Fang's expression and IMD           }
		\label{tab7}
\end{table*}

\begin{figure*}[h]\label{fig7}
	\includegraphics[width=1\textwidth]{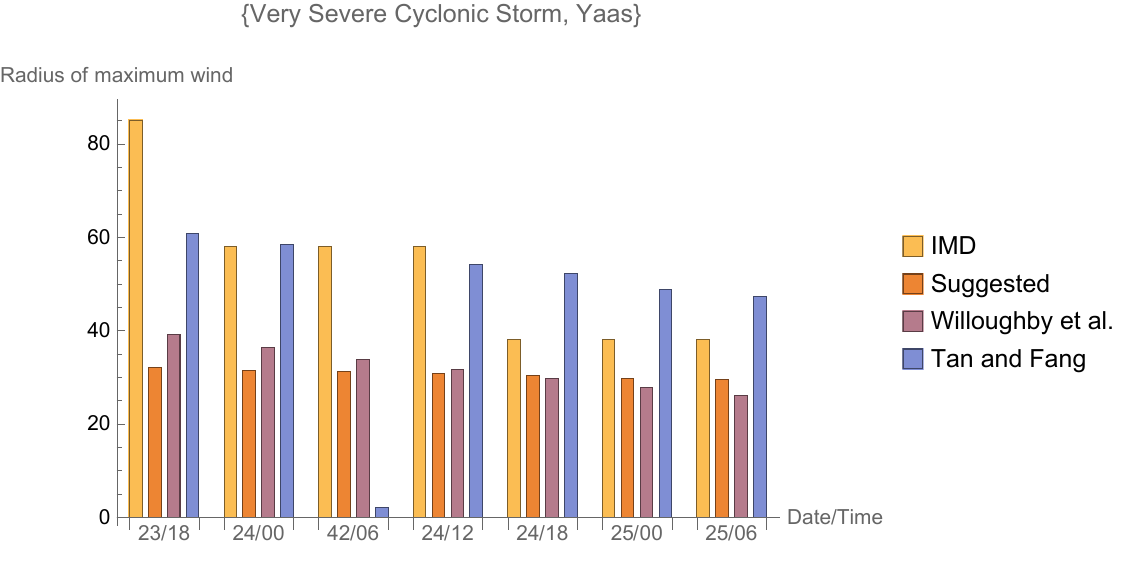}
	\caption{Very Severe Cyclonic Storm Yaas's value of $r_{max}$ given by suggested method, Willoughby et al., Tan and Fang, and IMD is graphically represented}
\end{figure*}

\section{Conclusion} \label{sec4}
This paper focuses on determining the value of $r_{max}$ across the NIO. There is a lack of literature specifically addressing $r_{max}$ in the NIO, despite it being an area surrounded by the Bay of Bengal, which is globally recognized as highly susceptible to TCs. Through an analysis of historical observations, we investigate the relationship between the position of the TC's center (latitude and longitude) and the estimated pressure drop at the center, in order to develop a relation for calculating $r_{max}$.

To evaluate the accuracy of our proposed approach, we employ two statistical methods namely Error percentage and T-test. By examining numerous TC cases within the North Indian basin, we find that both statistical methods favor our suggested approach in terms of accuracy compared to other methodologies \cite{bib21,bib22}. However, it is important to note a limitation of our method, which is that the value of $P_d$ must be less than or equal to 12 hPa, and there are no specific conditions imposed on the latitude of the TC center within the North Indian basin.

In summary, this study contributes to the understanding of $r_{max}$ in the NIO by conducting a comprehensive analysis of the relationship between TC center characteristics and estimated central pressure drop. The proposed approach demonstrates superior accuracy when compared to alternative methodologies, providing valuable insights for future research and forecasting of TCs in the NIO.

\section{Future Scope}\label{sec5}
Our future research focuses on exploring additional TC scenarios and developing a formula to estimate pressure drops exceeding 12 hPa at the center. In order to ascertain the value of $r_{max}$, we aim to validate this formula by examining additional basins, including the South Pacific and North Pacific basins. This extended investigation will provide a more comprehensive understanding of TC behavior and contribute to improved predictions of characteristics of TCs. By studying a wider range of basins, we can gather valuable data to refine our expression and enhance its accuracy. Consequently, this research will enhance our knowledge of TC dynamics and advance our ability to forecast pressure changes during these severe weather events.

\section*{Acknowledgement}
The best track data and metrics for TC characteristics are freely available thanks to the Indian Meteorological Department for providing the information at their website.
\section*{Declarations}
\subsection*{Conflict of Interest}
Conflict of interest is not declared by any of the authors.
\subsection*{Author Statement}
Research, data curing, and manuscript preparation was done by first author Monu Yadav. Research, Supervision, Review, and editing by Second author Laxminarayan Das
\subsection*{Data Availability}
The datasets include TC metrics such as TC number, time (year, month, day, and hour), locations (longitude and latitude of TC centre), estimated central pressure ($P_{c}$), maximum wind speed ($V_{m}$), and estimated pressure drop at the centre ($P_{d}$), which are stored in the datasets at a time interval of 3 hours is freely available on the website of Indian Meteorological Department.

\bibliography{sn-bibliography}


\end{document}